# Prospects for discovering new gauge bosons, extra dimensions and contact interaction at the LHC

P. Spagnolo
*INFN Pisa, ITALY*

This talk is a review of possible discoveries of exotic not Standard Model Physics in the early stage of LHC (first two years), with the ATLAS and CMS experiments. LHC will be the first experiment ever to explore the TeV region and new exotic particles could be discovered at early stages of the LHC running if their masses are not too large. This talk is focused on three main search topics: new vector bosons, contact interaction and extra dimensions.

## 1. INTRODUCTION

During the past decennia, many theories of new physics have been formulated such as supersymmetry, new strong forces, or theories with extra space-time dimensions, which go beyond the present Standard Model. It is very uncertain the way in which new physics would manifest itself. However, these possible new physics events are often expected with similar signatures: high invariant mass of the final state, isolated high-$p_T$ leptons, large missing $E_T$, very energetic photons and jets. Therefore it is possible that the first LHC data would not be enough to distinguish between the different models even if a clear signal of new physics will be observed.

The LHC is a proton-proton collider. The advantage of such hadron accelerators is that they allow for exploration of a wide range of energy/mass scales with fixed-energy beams: they are the natural choice for a discovery machine. This is because in hard proton-proton collisions the interaction actually occurs between their constituents (quarks and gluons), which carry a variable fraction of the proton's momentum. On the other hand, the LHC detectors will operate in a very difficult environment: the high bunch crossing frequency, the high event rate and the pile-up of several interactions in the same bunch crossing impose strict requirements on the detector design. The LHC detectors will also have to withstand an extremely high radiation dose and special radiation-hard electronics must be installed.

In these rather complicated hadronic events, the Drell-Yan production of lepton pairs is the process with the cleanest signature and lowest experimental background. Note that several of the past discoveries in Particle Physics have been achieved by observing an excess in the invariant mass of the Drell-Yan spectrum; the discoveries in 1974 of the J/$\psi$, in 1977 of the Y(4S) and in 1983 of the Z boson have all been obtained in the Drell-Yan channel. Therefore it is a natural strategy to search for these new particles as a possible excess in the dilepton spectrum at high mass. The advantage of this clean signature is that its background is well known theoretically and thus under control. It is given by the Standard Model Drell-Yan process with a rate above 1 TeV which is negligible compared to the expected signal. Besides, other simple and signatures, like e large missing $E_T$, energetic jets in the final state and $\gamma\gamma$ events can be used to claim a discovery. This is a review of possible early discoveries of new vector bosons, contact interaction and extra dimensions at LHC.

## 2. NEW VECTOR BOSONS

A large class of exotic models predicts the existence high-mass vector boson resonances in the TeV range of energy. Among these scenarios, there is a large variety of heavy neutral gauge bosons, named Z', predicted by many superstring-inspired and grand unification theories (GUT), as well as in dynamical symmetry breaking and "little Higgs" models [1]. LHC will be the first experiment ever to explore the TeV region and these new particles could be discovered at early stages (first two years) of the running of the LHC if their masses are not too large, not exceeding about 2 TeV. At the LHC all these gauge bosons can be observed as resonances decaying into two fermions, thus leading to a rather general signature of new physics. The golden channel for these searches is the dilepton decay, where two isolated leptons with high momentum and invariant mass are selected as a resonance in the Drell-Yan process $pp \rightarrow X \rightarrow l^+l^-$. In the following figure the diagram for the Standard Model Drell-Yan process is shown together with the mass peak of a hypothetical Z' resonance at 1 TeV compared to the expected background.

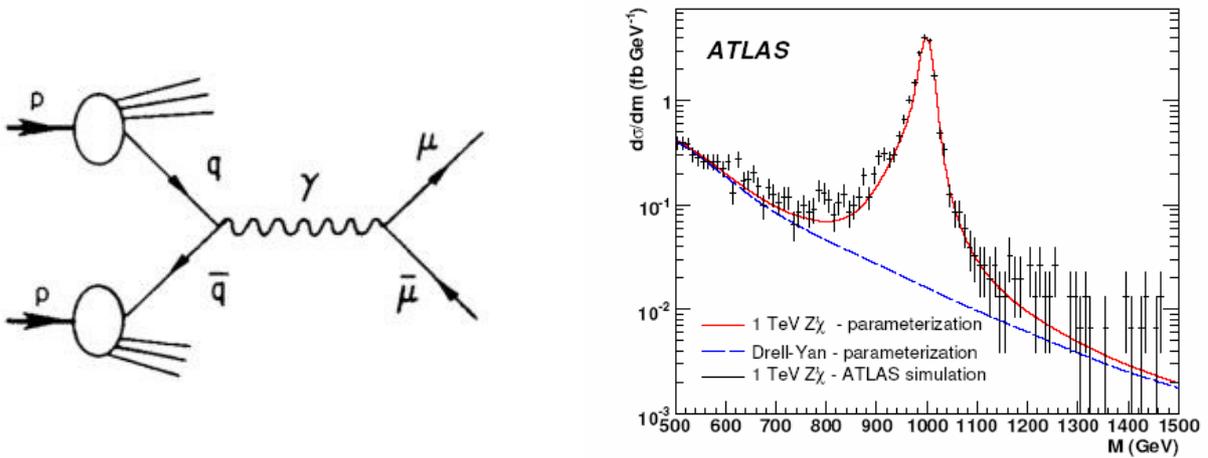

Figure 1: Drell-Yan di-muon process and expected mass shape for 1 TeV Z' decaying into two electrons in ATLAS.

LHC will allow to discover (or exclude) a Z' mass up to 5 TeV as shown in figure 2.a where the CMS discovery reach for dielectrons and dimuons is plotted as function of the luminosity for two different theoretical models. For these leptonic channels an excellent understanding of the detectors, the calibrations and the alignment are necessary. In figure 2.b the effect of the mis-alignment on the mass reconstruction in ATLAS is shown.

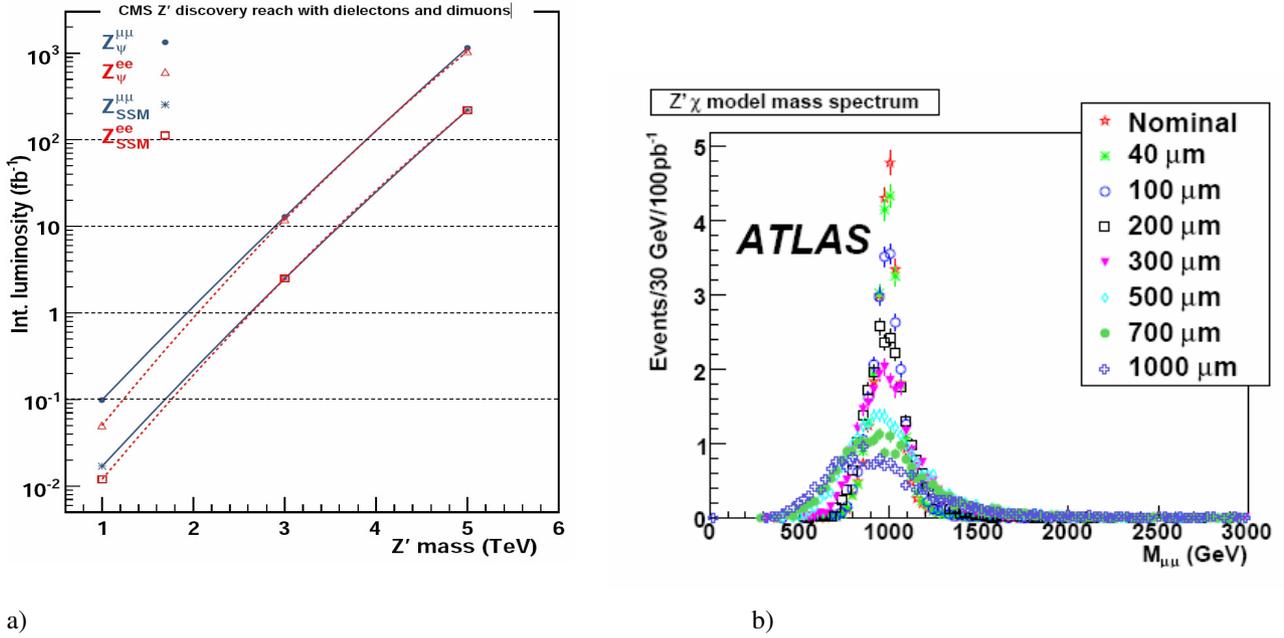

a)                                                                                         b)

Figure 2: a) Discovery potential of Z' in CMS for two different theoretical models. b) Mass resolution for Z' decay in dimuons for different mis-alignment scenarios in ATLAS

## 3. EXTRA-DIMENSIONS

Several extra dimensions theories allow observing at the TeV scale the graviton, the hypothetical elementary particle that mediates the force of gravity in the framework of quantum field theory. Randall-Sundrum (RS) models [2] predicts massive gravitons coupling both to fermions and bosons. At the LHC energies the RS graviton could be discovered in the dilepton final state, distinguishable from the Z' only by a different angular distribution in the decay due to the spin-2 of the graviton, or in the diphoton decay. Currently, no evidence of this new physics has been found and lower limits on the mass of these exotic particles are below 1 TeV. The discovery potential range of RS graviton in CMS is shown in figure 3.a. Besides, for a large extra-dimensions scenario, Arkani-Hamed Dimopoulos Dvali (ADD) models [3] predict gravitons at a scale $M_D$ related to the Planck scale $M_{Planck}$, the number of extra dimensions N and radius of the compactified space R by $M^2_{Planck} = M^{2+N}_D R^N$. The mass reach for the ADD graviton discovery limit is shown in figure 3.b. One of the possible consequences of large extra dimensions is the production of microscopic black holes at LHC when the collision impact parameters are below the Schwarzschild radius in 4+N dimensions, function of the reduced Planck mass. If $M_{PLANCK}$ is of the order of the TeV, the expected cross-section for such black holes production would be of the order of the pb. These black holes have a very short lifetime ($10^{-12}$ fs) and are expected to evaporate isotropically by the emission of all particles existing in nature. Potentially the BH observation could be accomplished with an integrated luminosity of 1 fb$^{-1}$ if $M_{PLANCK}$ is lower than 5 TeV.

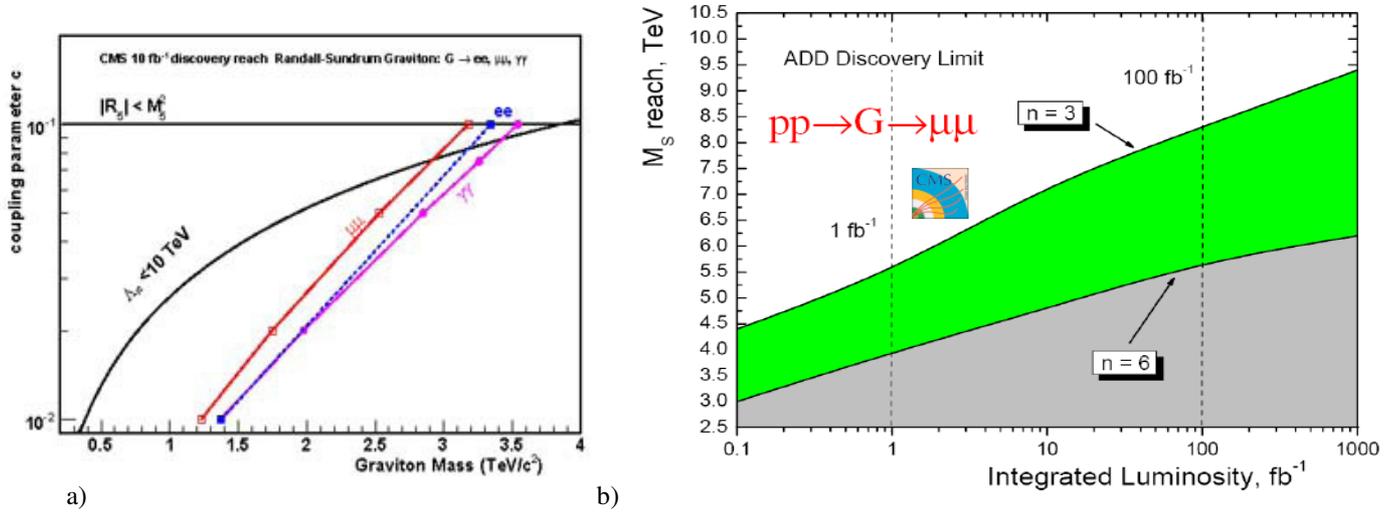

Figure 3: Discovery reach for Randall-Sundrum (a) and ADD (b) gravitons.

## 4. CONTACT INTERACTIONS

New physics at a scale $\Lambda$, which could be significantly larger than the available collision energy, can be modeled as contact interactions [4], with an effective four fermions vertex Lagrangian without a mediating vector boson. In the hadronic final states, contact interactions produce a rise in rate at high inclusive jet $p_T$, relative to QCD, as shown in figure 4. The error is dominated by the jet energy scale (~10%) in early running, but for jet with $p_T > 1$ TeV an immediate discovery is feasible, already with 10 pb$^{-1}$, beyond the Tevatron exclusion (up to $\Lambda \sim 3$ TeV ).

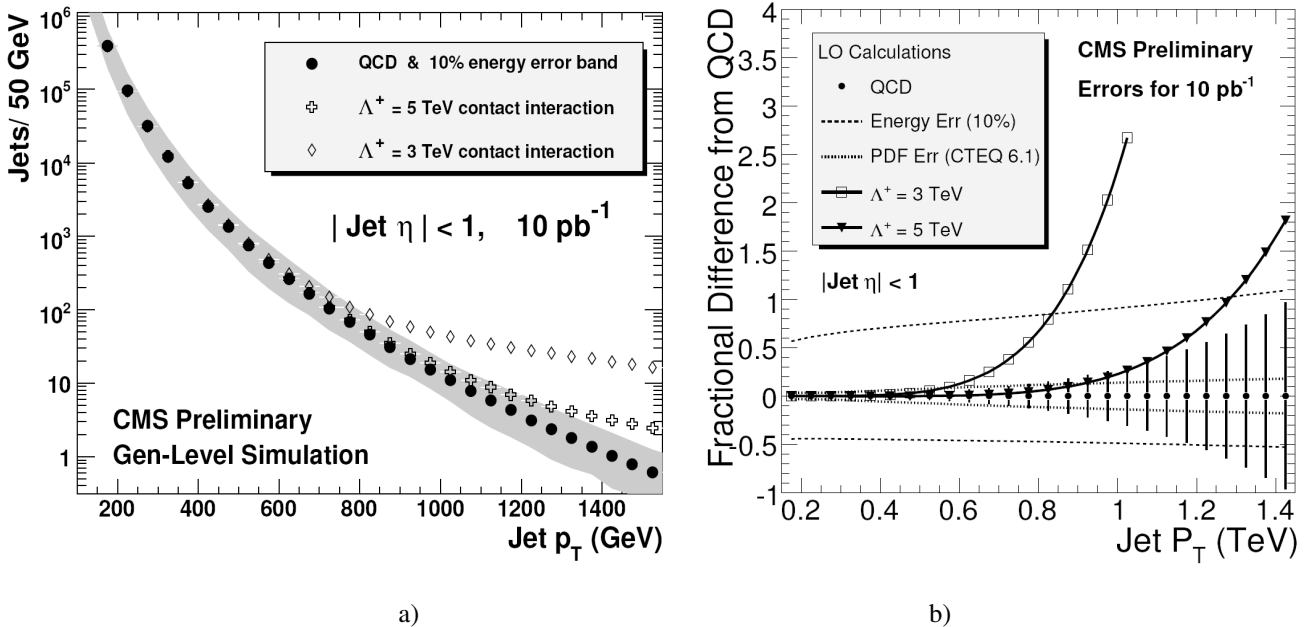

Figure 4: Excess in the Jet $P_T$ spectrum (a) and fractional difference respect to QCD (b) as function of jet momentum for a contact interaction of $\Lambda = 3$ (5) TeV.

Contact interactions can be discovered also in the dimuon final state, as an excess over the dimuon mass spectrum.

At the end of the LHC program, with 100 fb$^{-1}$, contact interactions up to Λ ~ 20 TeV could be easily discovered.

## 5. CONCLUSIONS

ATLAS and CMS will have the possibility to discover new physics beyond the Standard Model already with the early stage of LHC data . In Table 1 a summary of such possible discoveries are reported, together with the main sources of systematics that will affect the first data taking.

Table 1: Summary of the principal discoveries accessible with the first data taken at LHC.

| Model | Mass Reach (TeV) | L (pb$^{-1}$) | Early Systematic |
|---|---|---|---|
| Contact interaction | Λ ~ 2.8 | 10 | Jet efficiency and energy scale |
| **Z'** | | | |
| ALRM | M~1 | 10 | Alignment |
| SSM | M~1 | 20 | |
| LRM | M~1 | 30 | |
| E$_6$, SO(10) | M~1 | 30-100 | |
| Technirho | M~[0.3] | 100 | Jet energy scale |
| Axigluon or Colouron | M~[0.7,3.5] | 100 | Jet energy scale |
| Excited quark | M~[0.7,3.6] | 100 | Jet energy scale |
| E$_6$ di-quarks | M~[0.7,4] | 100 | Jet energy scale |
| mUED | M~ 0.3 − 0.6 | 10-1000 | MET, jet/photon energy scale |
| ADD real $G_{KK}$ | M$_D$ ~ 1.5 (n=3), ~ 1 (n=6) | 100 | MET, jet/photon energy scale |
| ADD virtual $G_{KK}$ | M$_D$ ~ 4.3 (n=3),~ 3 (n=6) | 100 | Alignment |
| | M$_D$ ~ 5 (n=3),~ 4 (n=6) | 1000 | |
| **RS1** | | | |
| di-jets | M$_G$ ~[0.7,0.8], c=0.1 | 100 | Jet energy scale, alignment |
| di-muons | M~[0.8,2.3], c=[0.01,0.1] | 1000 | |

## References

[1]  A. Leike, "The Phenomenology of Extra Neutral Gauge Bosons", Phys. Rep. 317 (1999) 143

[2]  L. Randall and R. Sundrum, Large Mass Hierarchy from a Small Extra Dimension", Phys. Rev. Lett. 83 (1999) 3370

[3]  N. Arkani, S. Dimopoulos and G.R. Divali, "The hierarchy problem and new dimension at a millimiter", Phys. Lett. B429 (1998) 263

[4]  S. Giddings and S. Thomas "High energy colliders as black hole factories: The end of short distance physics", Phys. Rev. D 65, 056010  (2002)

[5]  K.D. Lane, "Electroweak and flavour dynamics at hadron colliders" arXiv:hep-ph/9605257v1